\documentclass[12pt,english,twoside]{article}

\usepackage[T1]{fontenc}
\usepackage[utf8]{inputenc}
\usepackage[letterpaper, margin=2cm]{geometry}
\usepackage{color}
\usepackage{graphicx}
\usepackage{subfigure}
\graphicspath{{graphics_fold/}}
\usepackage{epstopdf} 
\usepackage{times}
\usepackage{etoolbox}
\usepackage{amsmath} 
\usepackage{amssymb}  
\usepackage{fancyhdr}
\usepackage{psfrag}
\usepackage{enumitem}
\usepackage{indentfirst}       
\usepackage{titlesec}          
\usepackage{hyperref}
\usepackage{babel}
\usepackage{listings}
\usepackage{booktabs}

\addto\captionsenglish{}

\usepackage[colorinlistoftodos]{todonotes} 

\setlist{nolistsep}

\makeatletter
\renewcommand*{\@biblabel}[1]{\hfill#1.}
\makeatother

\makeatletter
\patchcmd{\headrule}{\hrule}{\color{blue}\hrule}{}{}
\patchcmd{\footrule}{\hrule}{\color{blue}\hrule}{}{}
\makeatother

\fancypagestyle{firstpage}{
\fancyhead{}
\fancyfoot{}
\fancyfoot[LE,RO]{\small \thepage}

}

\makeatletter
\def\maketitle{
  \thispagestyle{firstpage}
\vspace*{-11mm}{\centering\includegraphics[width=0.86\textwidth]{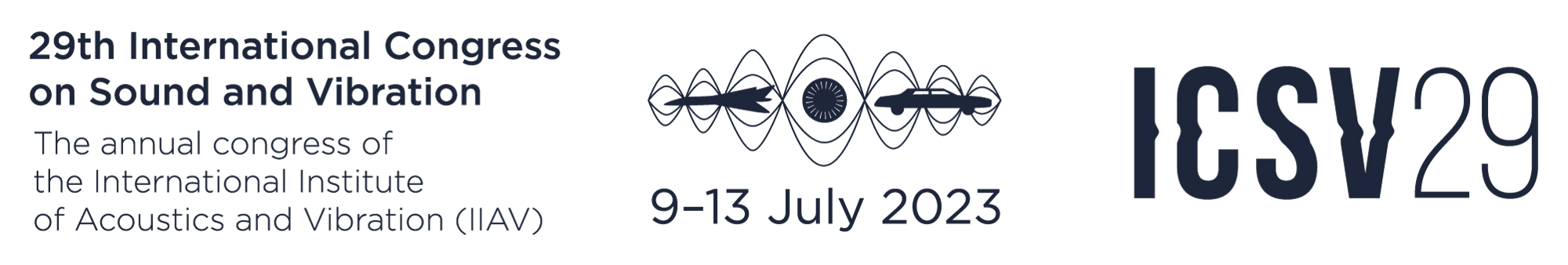}\\}
  {
   \vspace*{4mm}\fontsize{17}{20}\selectfont\sffamily{}  \noindent \MakeUppercase{\textbf{\@title}}

   \vspace*{3mm}\fontsize{14}{20}\selectfont\rmfamily{} \noindent \@author
  }
}
\makeatother

\titleformat{\section}
  {\fontsize{14}{14}\normalfont\sffamily\bfseries}
  {\thesection.}{1em}{}                

\titleformat{\subsection}
  {\normalfont\sffamily\bfseries}
  {\thesubsection}{1em}{}

\titleformat{\subsubsection}
  {\normalfont\sffamily\small}
  {\textit{\thesubsubsection}}{1em}\textit{{}}

\usepackage{mathrsfs}

\title{ADJOINT-BASED OPTIMIZATION OF THE CONTROL \\ BOUNDARIES OF A MICROFLUIDIC ACOUSTIC FLOW}

\author{Javier Lorente-Macias\\
{\small \textit{University of Cambridge, Cambridge, United Kingdom\\
e-mail: jl2269@cam.ac.uk}}\\
Matthew P. Juniper\\
{\small \textit{University of Cambridge, Cambridge, United Kingdom\\
email: mpj1001@cam.ac.uk}}}

\begin{document}

\maketitle
\renewcommand{\abstractname}{\vspace{-\baselineskip}} 

\begin{abstract}	\noindent

In this study, we find the optimal control boundary (i.e., actuator velocity) that cancels the acoustic reverberations inside drop-on-demand inkjet printheads at a specific time. We formulate an optimization problem to minimize the total energy of the oscillating flow at a given time (i.e., the acoustic energy inside the microchannel and the surface energy of the droplet). We use the adjoint method to compute the gradient of the cost function with respect to the control boundary, and a gradient-based optimization algorithm to converge to the optimal solution. This methodology has been successfully applied to two generic inkjet printhead mechanisms: thin-film and bulk types. In both cases, the actuator first reduces the surface energy of the system by extracting fluid from the nozzle. In this process, acoustic waves also propagate through the channel and reverberate at the ends, which increases the acoustic energy of the system. The actuator then sends additional acoustic waves that cancel these reverberations. Both mechanisms have been able to reduce the total energy of the system by a factor of over 100 in comparison with the uncontrolled cases.

\noindent Keywords: microfluidics, piezoacoustics, actuation design, inkjet printing
\end{abstract}

\quad\rule{425pt}{0.4pt}

\section{Introduction}

Drop-on-demand (DOD) inkjet printing is a technique in which ink droplets are generated on demand and jetted towards a moving substrate. An inkjet printhead typically contains several hundred microchannels and nozzles. Ink flows continuously through the channels at constant pressure difference (between one side and the other) and at constant pressure (relative to the atmosphere) just above the nozzle. An increase of the pressure must be generated inside the channel to expel droplets. In piezoelectric DOD printing, an actuator placed on one or more sides of the channel moves a few tens of nanometers, which generates acoustic pressure waves. These waves eventually reach the nozzle and a droplet is expelled. The piezoelectric actuators are made of ceramic materials that exhibit the inverse piezoelectric effect (i.e., they deform when a voltage is applied).

Manufacturers seek to increase the jetting frequency while maintaining uniformity in droplet size and shape. The former objective speeds up the printing process, whereas the latter makes it more accurate, which is important in high precision applications (e.g. printing microcircuits \cite{tomaszewski2017drops}; manufacturing of passive electronic components such as capacitors \cite{chiolerio2014inkjet}, biomedicine \cite{Lorber_2013}, or advanced manufacturing \cite{hoath2016fundamentals}). As the time between droplet ejections decreases, the droplet shape and size becomes increasingly affected by the residual oscillations from the previous droplet ejection. Therefore, it is important to damp these reverberations within the desired time interval. 

In this study, we design the open-loop control for the actuator that damps the acoustic reverberations inside the microchannels within a given time interval. This active control strategy is easier to implement on already existing printheads than passive control strategies (e.g., geometric variations). We consider two simplified geometries of two inkjet printing strategies already employed in the industry.

\section{Problem formulation}
\subsection{Governing equations}

We consider the flow inside the inkjet printhead microchannel. We derive the governing equations from a low Mach number asymptotic expansion \cite{kungurtsev_juniper_2019, Kungurtsev2023}. We further consider that variations in temperature are negligible, and that the ratio of specific heats is $\gamma_{th}=1$. Under these assumptions, we can express the nondimensionalized momentum and energy equations of the oscillating flow as:

\begin{subequations}
\label{ThermoViscousAcousticFlow}
    \begin{align}
    & \frac{\partial \Tilde{\boldsymbol{u}}}{\partial t} = -\nabla \Tilde{p} + \frac{1}{\Tilde{Re}}\nabla\cdot \Tilde{\boldsymbol{\tau}} \label{ThermoViscousAcousticFlowEq1}
    \\
    & \frac{\partial \Tilde{p}}{\partial t} + \nabla \cdot \Tilde{\boldsymbol{u}} = 0  \label{ThermoViscousAcousticFlowEq2}
    \end{align}
\end{subequations}
where $\Tilde{\boldsymbol{u}}$ is the acoustic velocity, $\Tilde{p}$ is the acoustic pressure, $\Tilde{Re}$ is the Reynolds number of the acoustic flow, and $\Tilde{\boldsymbol{\tau}} = \nabla\Tilde{\boldsymbol{u}} + \left(\nabla\Tilde{\boldsymbol{u}}\right)^{T} - \frac{2}{3}\nabla\cdot\Tilde{\boldsymbol{u}}\mathbb{I}$ is the viscous stress tensor.

These equations have been nondimensionalized using the length of the channel, $L_{c}$, as the characteristic length, the speed of the sound, $c_{0}$, as the characteristic speed, and $p_{c}=\rho c_{0}^{2}$ as the characteristic pressure. Therefore, the Reynolds number associated with the oscillating flow is $\Tilde{Re}=\frac{\rho c_{0} L_{c} }{\mu}$, with $\rho$ and $\mu$ being, respectively, the density and the dynamic viscosity of the fluid. This allows us to define the time scale $t_{c}=\frac{L_{c}}{c_{0}}$, which is the time required for a wave to propagate through the entire length of the channel.

We now consider the reduced-order model for the flow inside the droplet described in \cite{Kungurtsev2023}. The meniscus is a spherical cap containing incompressible fluid. Thus the stress at the nozzle outlet is equal to the surface tension, $\Tilde{\boldsymbol{\sigma}} \cdot \boldsymbol{n}|_{\Gamma_{N}} = -\gamma_{s} \kappa \boldsymbol{n}$, where $\Tilde{\boldsymbol{\sigma}} = -p\mathbb{I} + \frac{1}{\Tilde{Re}}\Tilde{\boldsymbol{\tau}}$ is the total stress tensor, $\gamma_{s}$ is the nondimensional surface tension coefficient and  $\kappa$ is the nondimensional curvature. This approach decouples the flow inside the droplet from the flow inside the channel, so that we just need to simulate the flow inside the microchannel and consider the previous impedance boundary condition at the nozzle outlet boundary.

The volume of the droplet, $\Omega$, varies according to the amount of mass flux through the nozzle outlet boundary:

\begin{equation}
    \frac{d\Omega}{dt} = -\epsilon\int_{\Gamma_{N}}\Tilde{\boldsymbol{u}}\cdot\boldsymbol{n} d\Gamma
\end{equation}

For a spherical cap, the volume of fluid inside the meniscus depends on the curvature through the following expression:

\begin{equation}
    \Omega\left(\kappa\right) = \frac{\pi}{3}R_{n}^{3}\frac{1}{\kappa^{3} }\left(2+\sqrt{1-\kappa^{2}}\right)\left(1-\sqrt{1-\kappa^{2}}\right)^{2}
\end{equation}

This leads to an ordinary differential equation for the evolution of the curvature:

\begin{equation}
    \frac{d\Omega\left(\kappa\right)}{d\kappa}\frac{d\kappa}{dt} = -\epsilon\int_{\Gamma_{N}}\Tilde{\boldsymbol{u}}\cdot\boldsymbol{n} d\Gamma
\end{equation}

For the other boundaries, we consider homogeneous Dirichlet boundary conditions at the walls, $\Tilde{\boldsymbol{u}}_{\Gamma_{W}} = 0$, nonhomogeneous Dirichlet boundary conditions at the actuators, $\Tilde{\boldsymbol{u}}_{\Gamma_{Act}}=\mathcal{U}_{Act}\left(t\right)$, and stress free boundary conditions at the channel inlet/outlets $\Tilde{\boldsymbol{\sigma}}\cdot\boldsymbol{n}|_{\Gamma_{In/Out}} = 0$.

In this paper, the initial acoustic state of the flow is taken to be zero acoustic velocity, $\Tilde{\boldsymbol{u}}\left(t=0\right) = 0$, and zero acoustic pressure, $\Tilde{p}\left(t=0\right) = 0$. The initial meniscus curvature has a nonzero initial value, $\kappa\left(t=0\right) = \kappa_{0}$. We define the acoustic energy, $\mathscr{E}_{ac}\left(t\right)$, and the surface energy, $\mathscr{E}_{s}\left(t\right)$, as follows:

\begin{subequations}
\label{AcousticSurfaceEnergy}
    \begin{align}
    \mathscr{E}_{ac}\left(t\right) &= \frac{1}{2}\left(\Tilde{u}^{2} + \Tilde{p}^{2} \right)
    \label{AcousticEnergy}
    \\
    \mathscr{E}_{s}\left(t\right) &= \frac{1}{2}\frac{\gamma_{s}}{\epsilon}R_{n}\left|\Gamma_{n} \right|
    \label{SurfaceEnergy}
    \end{align}
\end{subequations}
where $\gamma_{s}$ is the surface tension, $R_{n}$, is the nondimensional nozzle radius, and $\Gamma_{n} = 2\pi R_{n}^{2}\frac{1}{\kappa^{2} }\left(1-\sqrt{1-\kappa^{2}}\right)$ is the surface area of the spherical cap.

In order to ensure droplet uniformity, the acoustic energy inside the channel and the surface energy of the spherical cap at the nozzle outlet must be the same before the next droplet is ejected. For simplicity, we set these energies to zero in this paper, but a non-zero curvature could be targeted if desired. Therefore, we define the cost function as the total energy (i.e., sum of acoustic energy plus surface energy) at a given time $T$:

\begin{equation}
\label{CostFunction}
    \mathscr{J} = \mathscr{E}_{ac}\left(T\right) + \mathscr{E}_{s}\left(T\right)
\end{equation}

In order to minimize (\ref{CostFunction}), we use the actuator velocity as the control. Therefore, we first propose a parametrization of the function $\mathcal{U}_{Act}$. The simplest way to parametrize this funtion in time is to use a parameter at every time step. However, this leads to a very large set of control parameters, and increases the dimension of the search space accordingly. This can cause there to be many local minima, and the optimization algorithm may not be able to find the global minimum. In order to regularize the problem, we reduce the dimensions of the search space by expressing $U_{Act}$ through a truncated Fourier series expansion:

\begin{equation}
    \mathcal{U}_{Act}\left(t\right) = \sum_{n=0}^{N}a_{n}\cos{\left(\frac{n2\pi t}{T}\right)} + \sum_{i=0}^{N}b_{n}\sin{\left(\frac{n2\pi t}{T}\right)}
\end{equation}

We then define our control vector $\boldsymbol{c}\in\mathbb{R}^{(N+1)\times 1}$ as follows:

\begin{equation}
    \boldsymbol{c} = \left[\begin{array}{cccccccccc}
         a_{0} & a_{1} & \cdots & a_{N} & | & b_{0} & b_{1} & \cdots & b_{N}
    \end{array}\right]^{T}
\end{equation}

We set the initial and final actuator velocity to zero, $\mathcal{U}_{Act}\left(t=0\right) = \mathcal{U}_{Act}\left(t=T\right) = 0$, which leads to the following constraint:

\begin{equation}
\label{ControlConstraint}
    \sum_{i=0}^{N}a_{n}=0
\end{equation}

Finally, we formulate the optimization problem as finding the optimal set of Fourier coefficients that minimize the functional (\ref{CostFunction}) subjected to the governing equations (\ref{ThermoViscousAcousticFlow}) and the control constraint (\ref{ControlConstraint}).

\subsection{Adjoint-based optimization}

We solve the optimization problem proposed in the previous section using a gradient-based algorithm (interior-point method implemented through the function \texttt{fmincon} in MATLAB). The finite differences approach requires as many solutions of the governing equations as number of control parameters that we consider. Thus, this method leads to long computational times for the cases considered in this work (i.e., unsteady three-dimensional PDEs and several tens of control parameters).

In this study, we consider the adjoint method for computing the gradients, which is a method whose computational cost is independent of the number of control parameters. The Lagrangian of the optimization problem stated previously is:

\begin{equation}
\begin{split}
    \mathscr{L} = \mathscr{J} &+ \left[\frac{\partial \Tilde{\boldsymbol{u}}}{\partial t}  +\nabla \Tilde{p} - \frac{1}{\Tilde{Re}}\nabla\cdot \Tilde{\boldsymbol{\tau}} , \Tilde{\boldsymbol{u}}^{\dagger}\right] + \left[\frac{\partial \Tilde{p}}{\partial t} + \nabla \cdot \Tilde{\boldsymbol{u}}, \Tilde{p}^{\dagger}\right] + \\ 
    &+ \left\langle \frac{d\Omega\left(\kappa\right)}{d\kappa}\frac{d\kappa}{dt}  +\epsilon\int_{\Gamma_{N}}\Tilde{\boldsymbol{u}}\cdot\boldsymbol{n} d\Gamma, \frac{\gamma_{s}}{\epsilon}\kappa^{\dagger} \right\rangle
\end{split}
\end{equation}
where $\left[\square,\square\right]=\int_{0}^{T}\int_{\Omega}\square^{T} \square d\Omega dt $, $\left\langle\square,\square\right\rangle = \int_{0}^{T}\square^{T}\square dt$, $\left\{\square,\square\right\} = \int_{\Omega}\square^{T} \square d\Omega $, and $\square^{\dagger}$ are the Lagrange multipliers, which are also known as the adjoint variables.

We now set the variations of the Lagrangian to zero. Grouping terms in $\delta\Tilde{\boldsymbol{u}}$ and $\delta \Tilde{p}$ leads to the adjoint equations:

\begin{subequations}
\label{AdjointThermoViscousAcousticFlow}
    \begin{align}
    & -\frac{\partial \Tilde{\boldsymbol{u}}^{\dagger}}{\partial t} = \nabla \Tilde{p}^{\dagger} + \frac{1}{\Tilde{Re}}\nabla\cdot \Tilde{\boldsymbol{\tau}}^{\dagger} \label{AdjointThermoViscousAcousticFlowEq1}
    \\
    & -\frac{\partial \Tilde{p}^{\dagger}}{\partial t} + \nabla \cdot \Tilde{\boldsymbol{u}}^{\dagger} = 0  \label{AdjointThermoViscousAcousticFlowEq2}
    \end{align}
\end{subequations}
where $\Tilde{\boldsymbol{u}}^{\dagger} $ is the adjoint acoustic velocity, $\Tilde{p}^{\dagger} $ is the adjoint acoustic pressure, and $\Tilde{\boldsymbol{\tau}}^{\dagger} = \nabla\Tilde{\boldsymbol{u}}^{\dagger} + \left(\nabla\Tilde{\boldsymbol{u}}^{\dagger} \right) - \frac{2}{3}\nabla\cdot\Tilde{\boldsymbol{u}}^{\dagger} $ is the adjoint viscous stress tensor.

Similarly, grouping terms in $\delta \kappa$, we obtain the adjoint counterpart of the curvature ODE:

\begin{equation}
    -\frac{d\Omega}{d\kappa}\frac{d\kappa^{\dagger} }{dt} = -\epsilon\int_{\Gamma_{N}}\Tilde{\boldsymbol{u}}^{\dagger}\cdot\boldsymbol{n} d\Gamma + \frac{d^{2}\Omega}{d\kappa^{2}}\frac{d\kappa}{dt}\kappa^{\dagger}
\end{equation}

The adjoint boundary conditions of the adjoint problem are $\Tilde{\boldsymbol{u}}_{\Gamma_{W}}^{\dagger}=0$, $\Tilde{\boldsymbol{u}}_{\Gamma_{Act}}^{\dagger}=0$, $\Tilde{\boldsymbol{\sigma}}^{\dagger}\cdot \boldsymbol{n}|_{\Gamma_{In,Out}}=0$ and $\Tilde{\boldsymbol{\sigma}}^{\dagger}\cdot \boldsymbol{n}|_{\Gamma_{N}}=-\gamma_{s}\kappa^{\dagger}\boldsymbol{n}$. The adjoint initial conditions are given by the final state of the direct variables, $\Tilde{\boldsymbol{u}}^{\dagger}\left(t=0 \right)=-\Tilde{\boldsymbol{u}}\left(t=T\right)$, $\Tilde{p}^{\dagger}\left(t=0 \right)=-\Tilde{p}\left(t=T\right)$ and $\kappa^{\dagger}\left(t=0 \right)=\kappa\left(t=T\right)$. Finally, the gradient of the cost function with respect to the control parameters is defined as:

\begin{equation}
\label{Gradient}
    \delta \mathscr{J} = \left\{ \Tilde{\boldsymbol{\sigma}}^{\dagger} \cdot \boldsymbol{n}, \delta \boldsymbol{u} \right\}_{\Gamma_{Act}}
\end{equation}
where $\Tilde{\boldsymbol{\sigma}}^{\dagger} = \Tilde{p}^{\dagger}\mathbb{I} + \frac{1}{\Tilde{Re}}\Tilde{\boldsymbol{\tau}}^{\dagger}$ is the adjoint stress tensor.

This result shows that the optimal actuator velocity is the one that makes the the integral of the adjoint stress along the control boundary equal to zero. Thus, each iteration of the optimization algorithm requires one solution of the direct problem (\ref{ThermoViscousAcousticFlow}), then one solution of the adjoint problem (\ref{AdjointThermoViscousAcousticFlow}), and finally the computation of the gradient as in (\ref{Gradient}). This process is repeated until a stopping criteria is satisfied.

\section{Applications}

\subsection{Simplified 3D Thin-film inkjet printhead}

\begin{figure}
\centering    
\includegraphics[width=\textwidth]{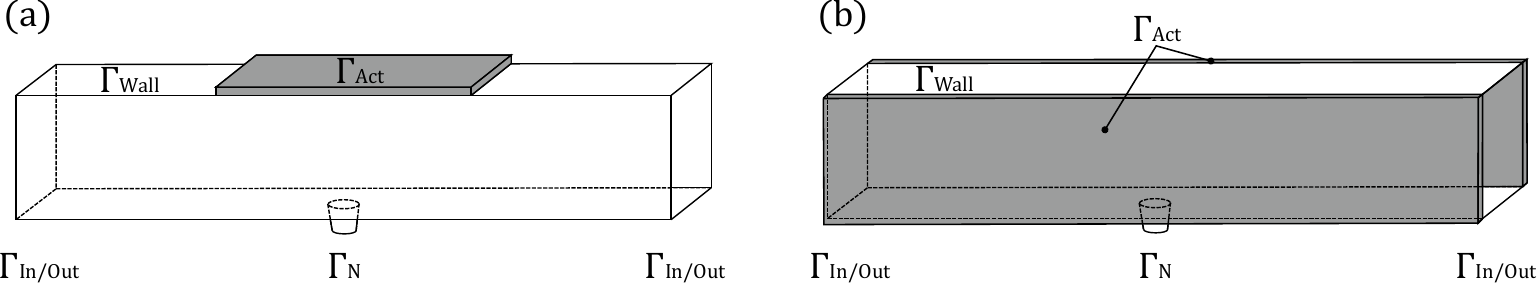}
\caption{Sketch of the geometries considered. (a) Thin-film type inkjet printhead; (b) Bulk type inkjet printhead. $\Gamma_{Wall}$ is the wall boundary, $\Gamma_{Act}$ is the actuator boundary, $\Gamma_{In/Out}$ is the inlet/outlet boundary, and $\Gamma_{N}$ is the nozzle outlet boundary.}
\label{Figure:Sketch}
\end{figure}

The first application of this optimization process is a simplified version of the thin-film inkjet printhead configuration (Fig. \ref{Figure:Sketch} a). This geometry consists of a three-dimensional microchannel of length $L_{c}$ with an actuator of length $L_{Act}=L_{c}/6$ placed on one of the faces, and a nozzle attached to the opposite face.

\begin{figure}
\centering    
\includegraphics[width=\textwidth]{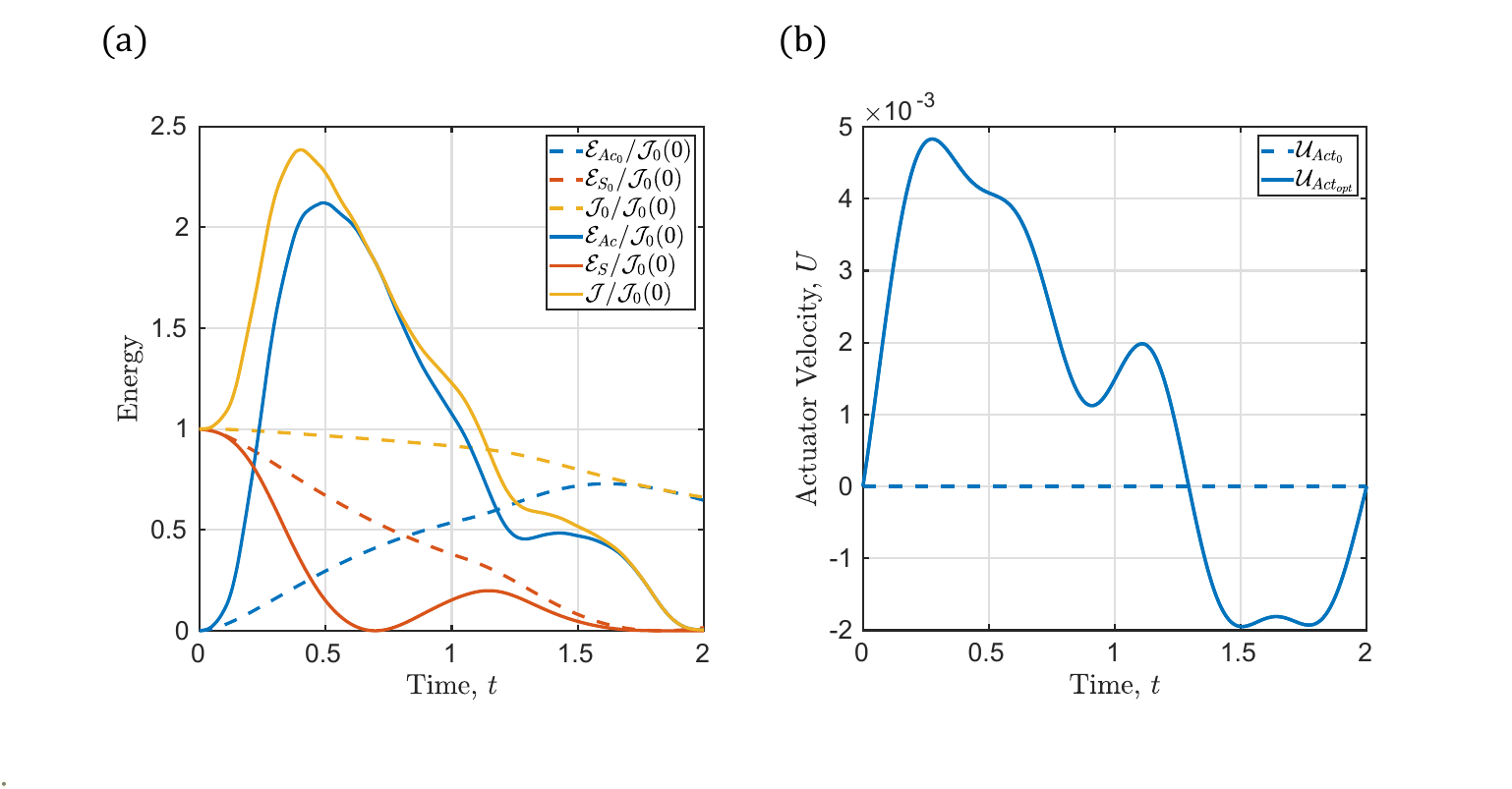}
\caption{Comparison of the uncontrolled (dashed lines) and controlled (solid lines) thin-film type inkjet printhead. (a) time evolution of the nondimensional cost functional and its terms (i.e., acoustic energy and surface energy); (b) actuator velocity.}
\label{Figure:1}
\end{figure}

Figure \ref{Figure:1} shows the time evolution of the acoustic and surface energy of the system considered for the uncontrolled and controlled cases. We observe that, without any actuation, there is a certain amount of acoustic energy remaining at $t=T$. In the controlled case,  the actuator first moves upwards, emitting a negative acoustic pressure wave. This wave is able to absorb the spherical cap, but also increases the reverberations within the channel. Thus, the acoustic energy during this first stage increases. At a second stage, the actuator moves downwards, cancelling the reverberations coming from the channel ends and driving the surface curvature back to zero. The final total energy of the controlled case is about 170 times smaller than in the uncontrolled case. The results and mechanism are similar to those found in a two-dimensional model by \cite{Kungurtsev2023}.


\begin{figure}
\centering    
\includegraphics[width=\textwidth, trim={0 4cm 0 4cm}]{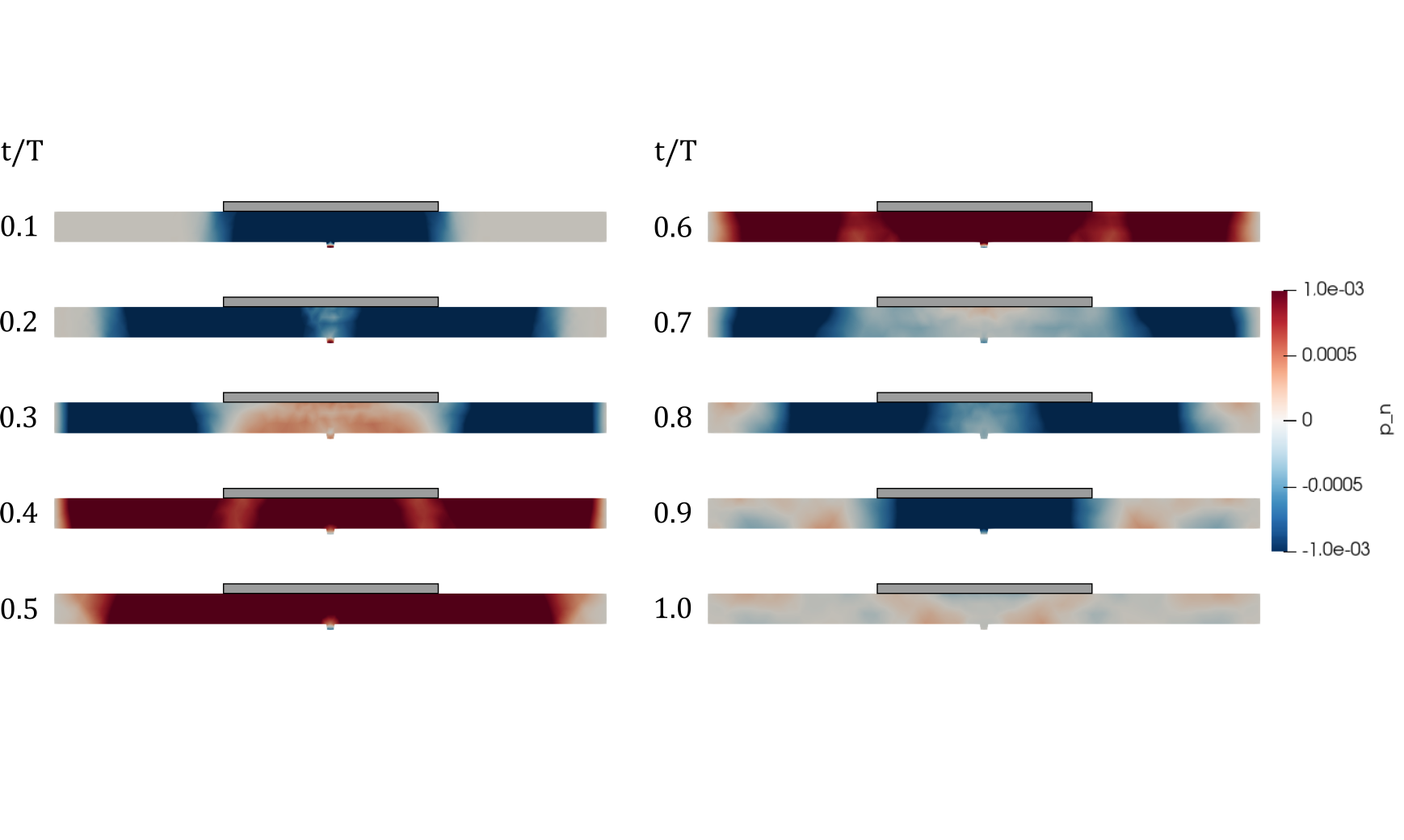}
\caption{Acoustic pressure field of the controlled thin-film type system at different times.}
\label{Figure:SnapshotsP4}
\end{figure}

Figure \ref{Figure:SnapshotsP4} shows ten snapshots of the acoustic pressure field inside the channel. We observe how the reverberations produced by the actuator propagate through the channel and are finally damped at $t=T$.

\subsection{Simplified 3D Bulk type inkjet printhead}

The second configuration has two actuators placed at two of the faces of the channel (Fig. \ref{Figure:Sketch} b). In this case, the actuator length is equal to the length of the channel, $L_{Act}=L_{c}$, and they move in an antisymmetric manner (i.e., $\mathcal{U}_{Act_{1}} = - \mathcal{U}_{Act_{2}}$).

\begin{figure}
\centering 
\includegraphics[width=\textwidth]{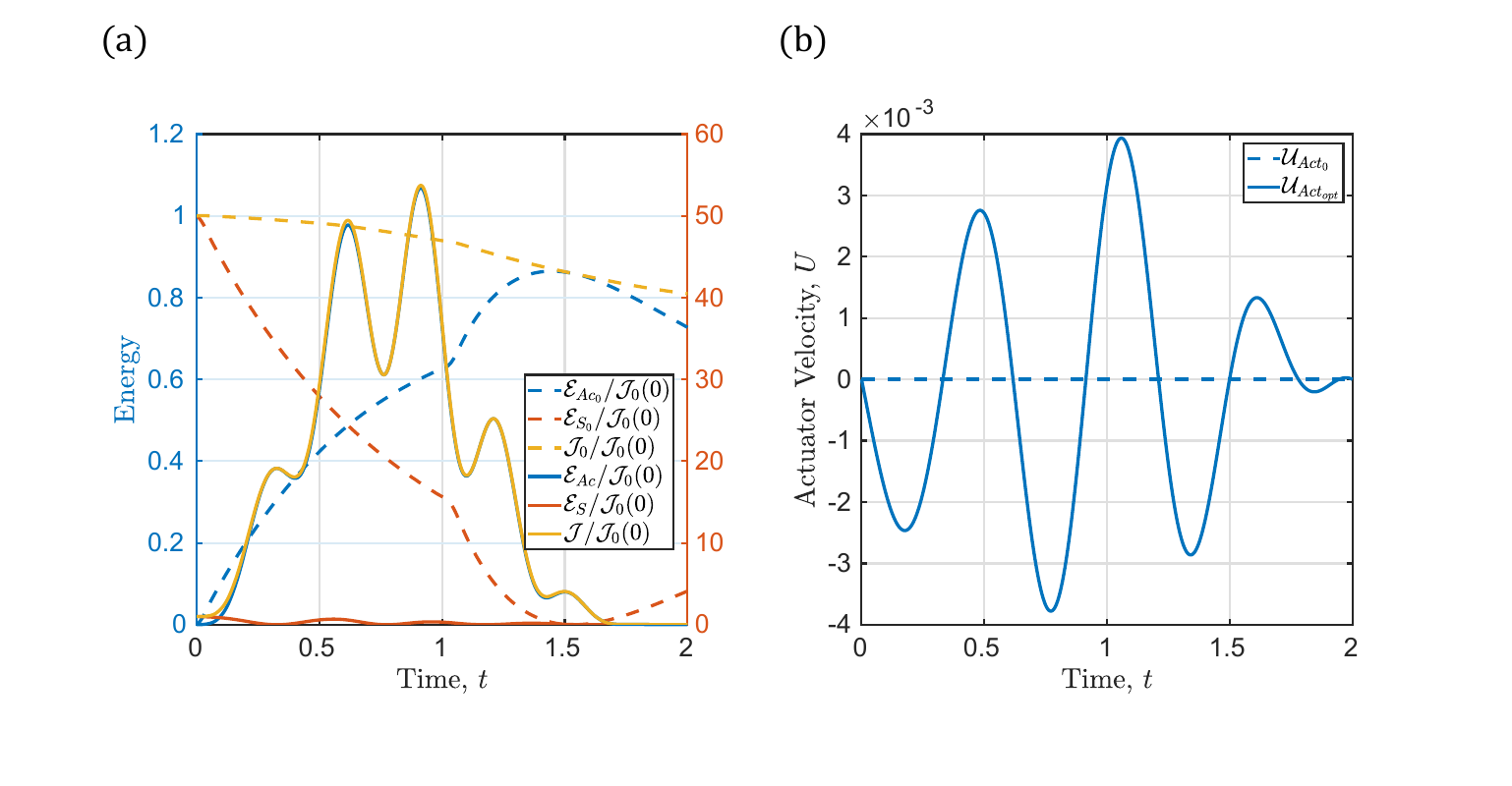}
\caption{Comparison of the uncontrolled (dashed lines) and controlled (solid lines) bulk type inkjet printhead. (a) time evolution of the nondimensional cost functional and its terms (i.e., acoustic energy and surface energy); (b) actuator velocity.}
\label{Figure:2}
\end{figure}

Figure \ref{Figure:2} shows the  time evolution of the cost function and the actuator velocity for the uncontrolled and controlled cases. In this case, the optimal open-loop control is more complicated, with the actuators moving inwards and outwards the channel several times to cancel the reverberations and turn the spherical cap into a flat surface. We can also observe how these motions increase and decrease the acoustic and surface energy during the time interval considered. The final energy of the controlled case is about 130 times smaller than in the uncontrolled case.

\begin{figure}
\centering    
\includegraphics[width=\textwidth, trim={0 4cm 0 4cm}]{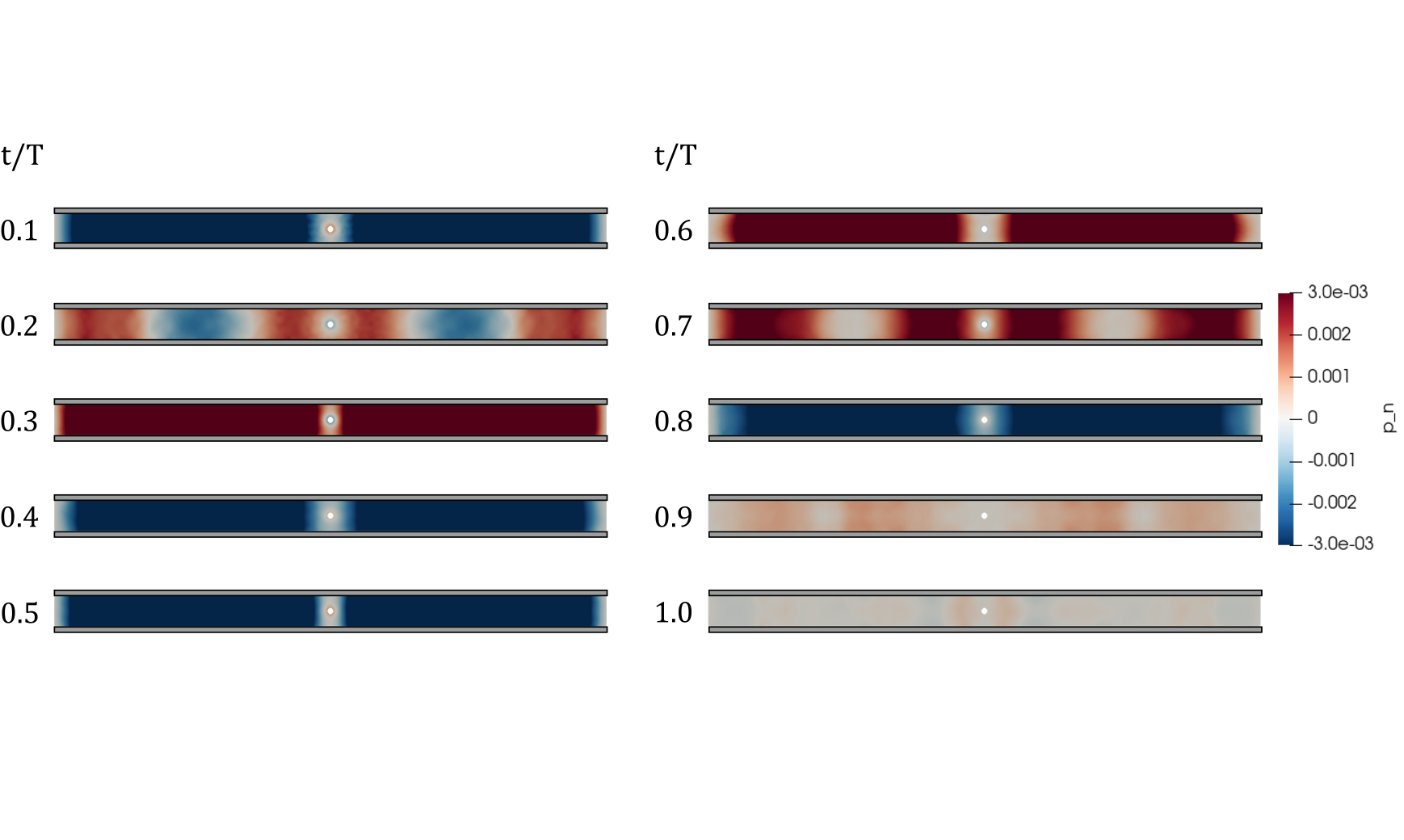}
\caption{Acoustic pressure field of the controlled bulk type system at different times.}
\label{Figure:SnapshotsP3}
\end{figure}

Figure \ref{Figure:SnapshotsP3} illustrates ten snapshots of the acoustic pressure field of the optimal case at different times. Again, we observe how the reverberations due to the actuator motion move through the channel, but they get damped at the final time $t=T$.

\section{Conclusion}

In the present study, we have applied an adjoint-based optimization approach to damp the residual reverberations that appear in piezoelectric DOD inkjet printheads. We have modelled the acoustic flow using a low Mach number asymptotic expansion, and have assumed that the droplet meniscus contains an incompressible flow. Under this assumption, the evolution of the droplet curvature depends on the mass flux through the nozzle outlet. The surface tension produced by this meniscus introduces an impedance boundary condition at the nozzle outlet, which decouples the flow inside the droplet from the one inside the channel.

Two control strategies that are currently being employed in the industry have been considered: thin-film and bulk type piezoelectric actuations. This method has allowed us to find the optimal open-loop control of the piezoelectric actuators for two simplified geometries. In both cases, the total energy of the flow (i.e., acoustic energy plus surface energy of the spherical cap) at the final time is significantly smaller than in the uncontrolled cases. The thin-film device is marginally easier to control, but both can be controlled.

Future research will focus on considering a better approximation of the complete inkjet printing chamber geometry (e.g., adding the ink manifolds manifolds or capturing the influence of the adjacent channels). The control laws are expected to change, but the implementation of the method is straightforward. Finally, the model proposed contains assumptions in its parameters. Therefore, we will assimilate experimental data using Bayesian inference to improve the accuracy of our physics based model and know their uncertainties.

\section*{Acknowledgements}

This project has received funding from the European Union’s Horizon 2020 research and innovation programme under grant agreement No 955923MARIE SKŁODOWSKA-CURIE ACTIONS Innovative Training Networks (ITN) Call: H2020-MSCA-ITN-2020.

\bibliographystyle{icsv_bib}
\bibliography{references}

\end{document}